\documentclass[11pt,a4paper]{article}
\usepackage[margin=2.5cm]{geometry}
\usepackage{amsmath,amssymb}
\usepackage{graphicx}
\usepackage{xcolor}
\usepackage{booktabs}
\usepackage{enumitem}
\usepackage[T1]{fontenc}
\usepackage{lmodern}
\usepackage[hyphens]{url}
\usepackage[colorlinks=true,linkcolor=blue!50!black,citecolor=blue!50!black,urlcolor=blue!50!black]{hyperref}
\usepackage{caption}
\setlist{itemsep=2pt,topsep=4pt}

\newcommand{\Ibar}{\bar I}
\newcommand{\ds}{\Delta_s}

\title{Formation and persistence of high-intensity ribbons\\
and dark regions in two-dimensional random media}

\author{M.~Billings\textsuperscript{1}, A.~Ji\textsuperscript{1},
D.~Milir\textsuperscript{2} and O.~Rath Spivack\textsuperscript{3,}\thanks{Corresponding author: \href{mailto:or100@cam.ac.uk}{or100@cam.ac.uk}}\\[6pt]
\small \textsuperscript{1}Faculty of Mathematics, University of Cambridge, UK\\
\small \textsuperscript{2}Mathematics Institute, University of Warwick, UK\\
\small \textsuperscript{3}Department of Applied Mathematics and Theoretical Physics,
University of Cambridge, UK}
\date{\today}

\begin{document}
\maketitle

\begin{abstract}
Waves propagating through a weakly scattering random medium form long, persistent
filaments of high intensity (``ribbons''), separated by persistent regions of low
intensity. An earlier study attributed the ribbons to large excursions of the
\emph{cumulative phase} of the medium, the phase accumulated along straight lines in
the direction of propagation. We revisit this explanation using the parabolic
equation in a two-dimensional continuous Gaussian medium, represented by independent
phase screens, with ensembles of 50 media. We define the statistics explicitly and
compare them with weak-fluctuation and saturated (Gaussian) theory. Up to about the
focusing range $x_f$, the intensity averaged over source position correlates
strongly with the transverse curvature of the cumulative phase (correlation
0.84--0.88), as geometric optics predicts; the correlation with the cumulative phase
itself is weaker near the source but declines more slowly, averaging about 0.4
between $3x_f$ and $6x_f$. It accounts for at most about half of the intensity variance, and scatter
plots of the original simulations show no consistent relation between the two beyond
the Fresnel distance. The transverse fourth moment of the intensity has a ribbon-like form only up
to about $2x_f$. Beyond that, for a plane wave, its height, width and
persistence length are those of Gaussian speckle; the width falls as $x^{-1/2}$, and
the persistence length is about $2.3\,k_0w^2$, where $w$ is the half-width.
Ribbon-scale structure survives mainly in the intensity averaged over source
position, whose correlation width stays near half the correlation length of the
medium. Moving the source by about half a correlation length decorrelates the
single-source pattern. We discuss the limits of the straight-line picture and the
statistical description of the persistent dark regions.
\end{abstract}

\section{Introduction}\label{sec:intro}

When a wave propagates through a random medium whose refractive-index fluctuations
are weak but smooth, its intensity does not simply become uniformly speckled. In
individual realisations it organises into long, narrow filaments of high intensity
which branch and persist over many correlation lengths, separated by extended
regions of low intensity. Such patterns have been observed and simulated in many
settings, from ocean acoustics and radio propagation to electron flow in
semiconductor heterostructures, and the phenomenon has become known as \emph{branched
flow}~\cite{Kaplan2002,Metzger2010}.

Uscinski and Spivack~\cite{US05} studied these ``ribbons'' in a two-dimensional
random medium using the parabolic equation, and averaged the intensity over the
position of the source to separate persistent, medium-induced structure from the
details of a single realisation. They proposed that a ribbon forms where the
\emph{cumulative phase} of the medium, the sum of the phase screens along a straight
line in the direction of propagation, has a large excursion. A large excursion tends
to be accompanied by large transverse curvature, so the medium acts locally as a
converging lens, and they argued that the ribbon persists because the scattering
angles remain small. The evidence was largely visual: a map of the product of the
source-averaged intensity and the cumulative phase. They also noted that some low
intensity regions remain dark as the source is moved, but did not analyse them.

The present paper revisits this picture quantitatively. It grew out of a summer
research project~\cite{BMJ} which re-implemented the simulations from the original paper, explored
alternative predictors of the ribbons, and computed fourth moments of the
intensity. We use a continuous Gaussian medium, so that the results do not depend on the numerical
grid, and ensembles of 50 media throughout. The results are compared with weak-fluctuation 
(geometric-optics) theory and with the Gaussian limit of saturated scattering. 
The aims are threefold: to test how
far, and over what range, the cumulative phase predicts the ribbons; to use the
fourth moment of the intensity to measure their transverse size and persistence;
and to begin a statistical description of the persistent dark regions.

Section~\ref{sec:model} describes the model and its numerical solution.
Section~\ref{sec:regimes} identifies the regimes of propagation through the
focusing range and the scintillation index. Section~\ref{sec:cumphase} examines the
correlation between cumulative phase and intensity, and Section~\ref{sec:m4} the
fourth moment, persistence and the effect of moving the source.
Section~\ref{sec:discussion} discusses the results and open questions.

\section{Model and numerical method}\label{sec:model}

\subsection{Parabolic equation}

We consider a time-harmonic scalar wave in two dimensions, with range $x$ and
transverse coordinate (height) $z$. In the paraxial approximation the reduced field
$E(x,z)$ satisfies the parabolic equation
\begin{equation}\label{eq:pe}
  \frac{\partial E}{\partial x}=\frac{i}{2k_0}\frac{\partial^2E}{\partial z^2}
  +\frac{ik_0}{2}\bigl(n^2-1\bigr)E,
\end{equation}
where $k_0=2\pi/\lambda$ and $n(x,z)$ is the refractive index. The intensity is
$I=|E|^2$.

\subsection{The random medium and its screen representation}\label{sec:medium}

The medium is $n=1+\mu h(x,z)$, where $h$ is a statistically homogeneous, isotropic
Gaussian random field with zero mean, unit variance and correlation function
$\exp\bigl(-(\xi^2+\zeta^2)/l^2\bigr)$. The strength is set by $\mu\ll1$ and the
correlation length by $l$.

We represent the medium by independent phase screens, as in the standard split-step
method. The range is divided into slabs of thickness $\ds$, and the slab
$[(i-1)\ds,\,i\ds]$ is replaced by a screen at its centre, which imposes the phase
it accumulates, $\phi_i(z)\approx k_0\mu\int h\,dx$. For $\ds\gtrsim l$, screens in
different slabs are effectively independent (the Markov approximation), and each
$\phi_i$ is Gaussian with transverse correlation $\exp(-\zeta^2/l^2)$ and variance
\begin{equation}\label{eq:sigs}
  \sigma_s^2=k_0^2\mu^2\sqrt\pi\,l\,\ds .
\end{equation}
The statistics of the field then depend on the medium only through $\mu$ and $l$,
not on $\ds$, provided $\ds$ is small compared with the ranges of interest.
Between screens the field propagates in free space, which is done exactly in the
Fourier domain. The free propagation is divided into sub-steps $\Delta x$ only to
record the field at intermediate ranges:
\begin{equation}\label{eq:step}
  E(x+\Delta x,\cdot)=\mathcal F^{-1}\Bigl[e^{-i\kappa^2\Delta x/2k_0}\,
  \mathcal F\bigl[E(x,\cdot)\bigr]\Bigr],
\end{equation}
with $E$ multiplied by $e^{i\phi_i(z)}$ on crossing screen $i$. Here $\kappa$ is the
transverse wavenumber and the $z$-domain is periodic. Each screen profile is
synthesised by filtering complex white noise with the square root of the Gaussian
spectrum, on the periodic $z$-grid.


Table~\ref{tab:conv} confirms that the statistics converge with respect to the screen 
spacing. Using independent screens with variance~\eqref{eq:sigs} for $\ds<l$ simply 
refines the discretisation of the delta-correlated (Markov) continuum limit, which is 
valid here because $l$ is small compared with $x_f$ and $x_F$. The scintillation index 
and the width of the intensity correlation (both defined in Section~\ref{sec:m4}) are 
unchanged, within the statistical error, for screen spacings from $0.2\,l$ to $2l$. 
We use $\ds=l$.

\begin{table}[ht]
\centering
\caption{Convergence with respect to screen spacing $\ds$ (plane wave, 20 media
each). SI is the scintillation index and $w$ the half-width at half-maximum of the
transverse intensity covariance~\eqref{eq:CT}. Standard errors are about 0.02 for SI
($\le0.01$ for $x\le25$) and 0.01 for $w$.}
\label{tab:conv}
\small
\begin{tabular}{lccccc|ccc}
\toprule
& \multicolumn{5}{c|}{SI at $x=$} & \multicolumn{3}{c}{$w$ at $x=$}\\
$\ds$ & 10 & 25 & 50 & 100 & 200 & 50 & 100 & 200\\
\midrule
0.6 & 0.03 & 0.32 & 1.00 & 0.97 & 1.03 & 0.80 & 0.64 & 0.46\\
1.2 & 0.02 & 0.34 & 0.95 & 0.97 & 1.04 & 0.80 & 0.63 & 0.46\\
3.0 & 0.02 & 0.32 & 0.90 & 0.95 & 1.09 & 0.81 & 0.62 & 0.46\\
6.0 & 0.02 & 0.34 & 0.91 & 1.02 & 1.05 & 0.81 & 0.63 & 0.45\\
\bottomrule
\end{tabular}
\end{table}

\subsection{Sources and source averaging}

Following~\cite{US05}, each source is a Gaussian beam,
$E(0,z;z_s)=\exp\bigl(-(z-z_s)^2\bigr)$. Its half-width of~1 is a third of the
correlation length, and its Rayleigh range $k_0/2\approx6.3$ is short compared with
the propagation ranges, so beyond a few units of range it behaves much like a point
source. For one realisation of the medium, the intensity averaged over source
position is
\begin{equation}\label{eq:Ibar}
  \Ibar(x,z)=\frac{1}{N_s}\sum_{s=1}^{N_s}\bigl|E(x,z;z_s)\bigr|^2,\qquad
  z_s=-20,-19,\dots,19,\quad N_s=40 .
\end{equation}
Where a statistically homogeneous field is needed, we also use a unit plane wave,
$E(0,z)=1$. Ensemble averages over realisations of the medium are written
$\langle\cdot\rangle$.

\subsection{Parameters}

The parameters are listed in Table~\ref{tab:params}. All lengths are in the same
(arbitrary) unit, in which the wavelength is 0.5. Statistics involving the sources
are taken over the window $|z|<15$, inside the span of source positions, and up to
$x=200$, beyond which the beams from the outermost sources begin to reach the
periodic boundary. Plane-wave statistics use the whole $z$-domain and extend to
$x=400$.

\begin{table}[ht]
\centering
\caption{Parameters and derived scales.}
\label{tab:params}
\small
\begin{tabular}{ll}
\toprule
Quantity & Value\\
\midrule
Wavelength $\lambda$; wavenumber $k_0$ & 0.5; $4\pi\approx12.57$\\
Correlation length $l$; rms index fluctuation $\mu$ & 3; 0.01\\
Screen spacing $\ds$; rms screen phase $\sigma_s$ & 3; 0.50 rad\\
Height of domain; $N_z$; $\Delta z$ & 200; 1500; 0.133\\
Recording step $\Delta x$; maximum range & 0.2; 200 (sources), 400 (plane wave)\\
Media per ensemble & 50\\
Cumulative-phase variance $\sigma_f^2(x)=k_0^2\mu^2\sqrt\pi\,l\,x$ & $0.084\,x$\\
Fresnel distance $x_F=k_0l^2$ & 113\\
Focusing range $x_f$, equation~\eqref{eq:xf} & 33.6\\
\bottomrule
\end{tabular}
\end{table}

\section{Regimes of propagation}\label{sec:regimes}

\subsection{Scintillation index and the onset of focusing}

The scintillation index $\mathrm{SI}=\langle I^2\rangle/\langle I\rangle^2-1$ is the
normalised intensity variance. As the wave penetrates an extended random medium it
passes through three regimes. In the \emph{weak} regime the index is small and grows
with range: the intensity is the undisturbed value plus a small perturbation set by
the accumulated phase curvature. In the \emph{focusing} regime the curvature is
strong enough for ray tubes to collapse, caustics form, and the scintillation index rises to a
maximum, usually above~1. In the \emph{saturated} regime many scattered
contributions add with effectively random phases, the field tends to a circular
complex Gaussian, the intensity distribution becomes exponential, and the scintillation index
tends to~1~\cite{Uscinski1977,Ishimaru1978}.

A simple estimate locates the onset of focusing. For a plane wave, geometric optics
gives the intensity as the inverse of the ray-tube Jacobian $J$. While the
perturbations are small,
\begin{equation}\label{eq:jac}
  \frac{\delta I}{I}\approx-(J-1)=-\frac{1}{k_0}\sum_i (x-x_i)\,
  \frac{\partial^2\phi_i}{\partial z^2},
\end{equation}
where the sum is over the screens before range $x$. For the Gaussian correlation,
$\operatorname{var}(\partial^2\phi_i/\partial z^2)=12\,\sigma_s^2/l^4$. The screens
are independent and $\sum_i(x-x_i)^2\approx x^3/(3\ds)$, so
\begin{equation}\label{eq:SIweak}
  \mathrm{SI}\approx\operatorname{var}(J-1)\approx\frac{4\sqrt\pi\,\mu^2x^3}{l^3}.
\end{equation}
The index grows as $x^3$, and caustics become common when it reaches order one, at
the focusing range
\begin{equation}\label{eq:xf}
  x_f\approx l\,\bigl(4\sqrt\pi\,\mu^2\bigr)^{-1/3}\approx33.6 .
\end{equation}
This has the form $l\,\mu^{-2/3}$ of the first-caustic distance in the branched-flow
literature, $\xi\varepsilon^{-2/3}$ for a potential of relative strength
$\varepsilon$ and correlation length $\xi$~\cite{Kaplan2002,Metzger2010}. The
parabolic equation is a Schr\"odinger equation with range as time, so
$\varepsilon$ corresponds to $\mu$ and $\xi$ to $l$.

Diffraction enters through the Fresnel scale $\sqrt{x/k_0}$, the width of the first
Fresnel zone, below which transverse structure is smoothed. It equals the
correlation length at the Fresnel distance $x_F=k_0l^2\approx113$. For $x\ll x_F$
each correlation length of the medium acts as a refracting lens and geometric optics
applies; for $x\gg x_F$, diffraction averages the phase curvature over several
correlation lengths, and in the weak regime the growth of the index slows to roughly
linear. Caustics can therefore form only if $x_f$ is not large compared with $x_F$.
Here $x_f/x_F\approx0.3$, and the Fresnel scale at $x_f$ is about 1.6, so focusing
occurs while geometric optics still holds approximately.

\subsection{Results}

\begin{figure}[ht]
\centering
\includegraphics[width=\textwidth]{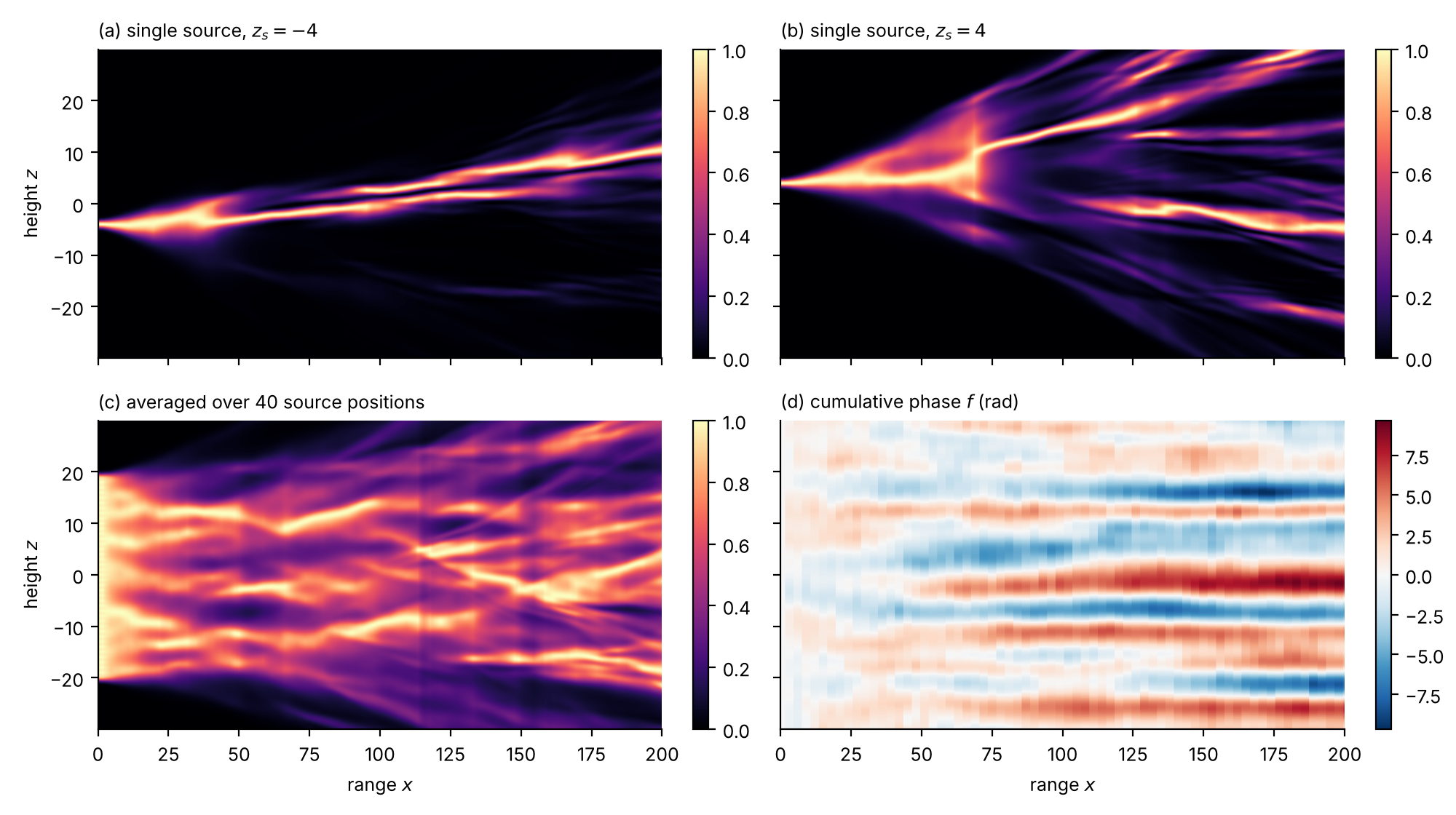}
\caption{One realisation of the medium. (a), (b) Single-source intensities for
sources at $z_s=-4$ and $z_s=4$; (c) intensity averaged over 40 source positions; in
(a)--(c) each range is normalised to its maximum. (d) Cumulative phase $f$ with its
mean over $z$ removed.}
\label{fig:fields}
\end{figure}

Figure~\ref{fig:fields} shows a typical realisation of the medium. Single-source fields (panels a, b)
show a branching pattern that changes markedly when the source moves by 8 units. The
source-averaged intensity (c) shows ribbons several units wide, slanting and
branching over tens of units of range, and dark bands between them. The cumulative
phase (d) is, by construction, a running sum in $x$, and its structures are close
to horizontal.

Figure~\ref{fig:plane}(a) in Section~\ref{sec:m4} shows the plane-wave scintillation index. It follows the
weak estimate~\eqref{eq:SIweak} closely up to $x\approx25$ (0.02 and 0.33 at
$x=10$ and 25, against 0.026 and 0.41), reaches about~1 by $x\approx50$
($1.5x_f$), and then stays between 1.0 and 1.1 out to $x=400$ ($12x_f$). The overshoot
is mild, the largest value being $1.12\pm0.02$ near $x=150$. A strong focusing peak
is characteristic of a single strong screen; in an extended medium the caustics of
successive regions overlap, and the index passes almost directly from weak
fluctuations to near-saturation. For comparison, the simulations of~\cite{US05}
(recovered from the stored codes and data) used screens of rms phase 0.139 at
spacing $0.01\,k_0L^2$, where $L$ is the correlation length, and extended to
$x=3k_0L^2$. In Fresnel units their medium was about five times weaker
($\sigma_f^2\approx1.9$ per $k_0L^2$, against 9.5 here). The plane-wave index for
media with the same statistics rises to about 0.8 by $x=k_0L^2$ and levels off just
below~1, again without a distinct peak.

\section{Cumulative phase and intensity}\label{sec:cumphase}

\subsection{Definitions}

Following~\cite{US05} we define the cumulative phase as the phase imposed by the medium alone,
 accumulated along
the horizontal line at height $z$:
\begin{equation}\label{eq:f}
  f(x,z)=\sum_{i:\,x_i\le x}\phi_i(z)\;\approx\;k_0\mu\int_0^x h(x',z)\,dx'.
\end{equation}
It is not the phase of the field. Its variance is $\sigma_f^2(x)=k_0^2\mu^2\sqrt\pi\,
l\,x$, so its rms reaches 1\,rad at $x\approx12$ and 2.9\,rad at $x=100$. (It should not
be reduced modulo $2\pi$: once its rms exceeds about 1\,rad the wrapped value no
longer measures the size of the excursion, and correlations computed from it decay
spuriously.) We remove its mean over $z$ at each range, which has no effect on the
intensity. Near the source the relevant quantity, by~\eqref{eq:jac}, is the
transverse curvature $-\partial^2 f/\partial z^2$, which is positive at a local
maximum of $f$, that is, at a converging lens. Formula~\eqref{eq:jac} in fact weights
the screens by their distance $x-x_i$ from the observation range, so $f$ and its
curvature are approximations to the quantity it involves.

The comparison uses the correlation coefficient (normalised covariance) over heights,
taken separately at each range. For fields $A$
and $B$ and a fixed range $x$,
\begin{equation}\label{eq:r}
  r_{AB}(x)=\frac{\sum_{m}\bigl(A(x,z_m)-\langle A\rangle_W\bigr)\bigl(B(x,z_m)-\langle B\rangle_W\bigr)}
  {\bigl[\sum_m\bigl(A-\langle A\rangle_W\bigr)^2\sum_m\bigl(B-\langle B\rangle_W\bigr)^2\bigr]^{1/2}},
\end{equation}
where the sums run over the grid points $z_m$ in the window $W=\{|z|<15\}$ (225
points, about 10 correlation lengths) and $\langle\cdot\rangle_W$ is the mean over
$W$. Taking the correlation one range at a time makes it insensitive to the decay of
$\Ibar$ with range and to any $z$-independent offset in $f$. We compute $r$ with
$A=\Ibar$ and $B=f$, $B=-\partial^2f/\partial z^2$, and, as a null, $B$ equal to the
cumulative phase of an independent medium. Correlations are averaged over bands of
range and then over the 50 realisations.

To separate bright from dark regions, we standardise $f$ over $W$ at each range,
$f^*=(f-\langle f\rangle_W)/\mathrm{sd}_W(f)$, and record the mean of $f^*$ over the
points where $\Ibar$ lies in its lowest or highest 10\% at that range, pooled over
each band.

For reference, the ``correlation'' map of~\cite{US05} is the pointwise product
$\Ibar f$, with neither mean removed nor normalisation; it shows where both
quantities are large, but is not a correlation coefficient.

\subsection{Results}

\begin{figure}[ht]
\centering
\includegraphics[width=0.85\textwidth]{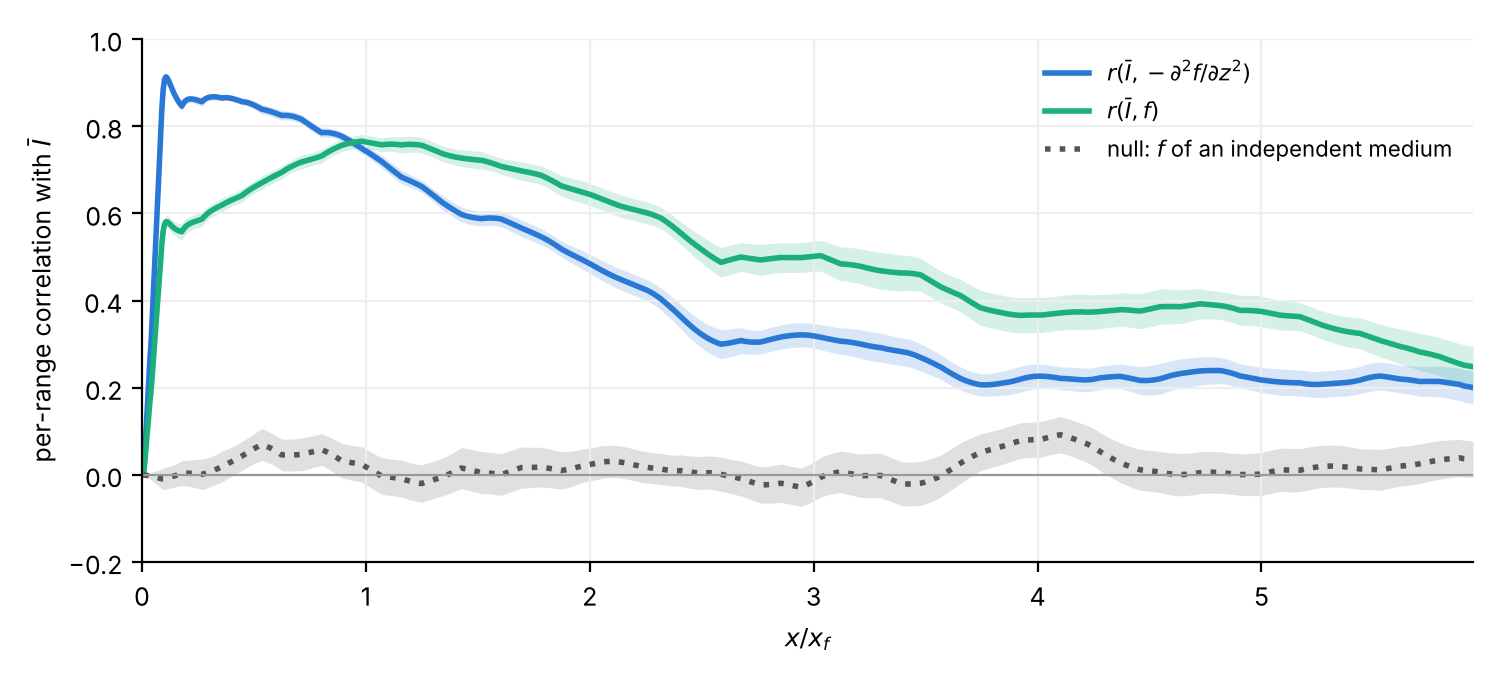}
\caption{Per-range correlation~\eqref{eq:r} of the source-averaged intensity with
the transverse curvature of the cumulative phase, with the cumulative phase itself,
and with the cumulative phase of an independent medium (null). Mean over 50 media,
shaded $\pm1$ standard error, smoothed over one screen spacing.}
\label{fig:corr}
\end{figure}

\begin{table}[ht]
\centering
\caption{Band-averaged correlations with $\Ibar$ and conditional means of the
standardised cumulative phase (50 media, mean $\pm$ s.d.\ over media; the standard
error of each mean is about a seventh of the s.d.).}
\label{tab:corr}
\small\setlength{\tabcolsep}{3.5pt}
\begin{tabular}{lccccc}
\toprule
Range & $r(\Ibar,-\partial^2f/\partial z^2)$ & $r(\Ibar,f)$ & null & $f^*$, darkest 10\% & $f^*$, brightest 10\%\\
\midrule
2--10    & $0.88\pm0.05$ & $0.58\pm0.10$ & $0.00\pm0.18$ & $-0.98\pm0.45$ & $+0.99\pm0.32$\\
10--25   & $0.84\pm0.05$ & $0.66\pm0.08$ & $0.04\pm0.23$ & $-1.12\pm0.41$ & $+1.16\pm0.33$\\
25--50   & $0.70\pm0.07$ & $0.75\pm0.10$ & $0.01\pm0.26$ & $-1.33\pm0.38$ & $+1.24\pm0.32$\\
50--100  & $0.43\pm0.11$ & $0.60\pm0.13$ & $0.01\pm0.21$ & $-1.03\pm0.29$ & $+0.98\pm0.28$\\
100--200 & $0.23\pm0.11$ & $0.38\pm0.16$ & $0.02\pm0.20$ & $-0.66\pm0.41$ & $+0.56\pm0.30$\\
\bottomrule
\end{tabular}
\end{table}

Figure~\ref{fig:corr} and Table~\ref{tab:corr} give the results. Four features stand
out.
\begin{enumerate}
\item \textbf{Near the source, curvature is the better predictor.} Up to about $x_f$
  the correlation of $\Ibar$ with $-\partial^2f/\partial z^2$ is 0.84--0.88, as the
  geometric-optics relation~\eqref{eq:jac} implies. Beyond $x_f$ it falls steadily, to
  about 0.2 at $6x_f$, as the linearisation fails.
\item \textbf{The cumulative phase itself correlates more weakly but more
  persistently.} Its correlation rises to about 0.75 near $x_f$ and declines slowly,
  averaging about 0.4 between $3x_f$ and $6x_f$ and reaching about 0.25 at $6x_f$. Part of this persistence is built in: $f$ is a running sum
  in $x$, so neighbouring ranges are far from independent, and a ribbon that has once
  formed over a maximum of $f$ remains correlated with it while it persists. 
  For a Gaussian correlation function, the correlation between $f$ and its curvature has magnitude only $1/\sqrt3\approx0.58$, so a large excursion of $f$ is only loosely a strong lens.
\item \textbf{Single realisations are unreliable.} The null has ensemble mean close to
  zero, but in a single medium its band average spreads by $\pm0.2$. A single-medium
  correlation of order 0.3 could therefore arise by chance; only ensemble averages are
  meaningful.
\item \textbf{Bright and dark regions are tied to the phase symmetrically.} The
  brightest 10\% of $\Ibar$ lie on average about one standard deviation above the mean
  of $f$, and the darkest 10\% about one below, up to $3x_f$. The association weakens
  at larger range, for both.
\end{enumerate}

\subsection{Scatter plots in the original simulations}\label{sec:scatter}

\begin{figure}[ht]
\centering
\includegraphics[width=\textwidth]{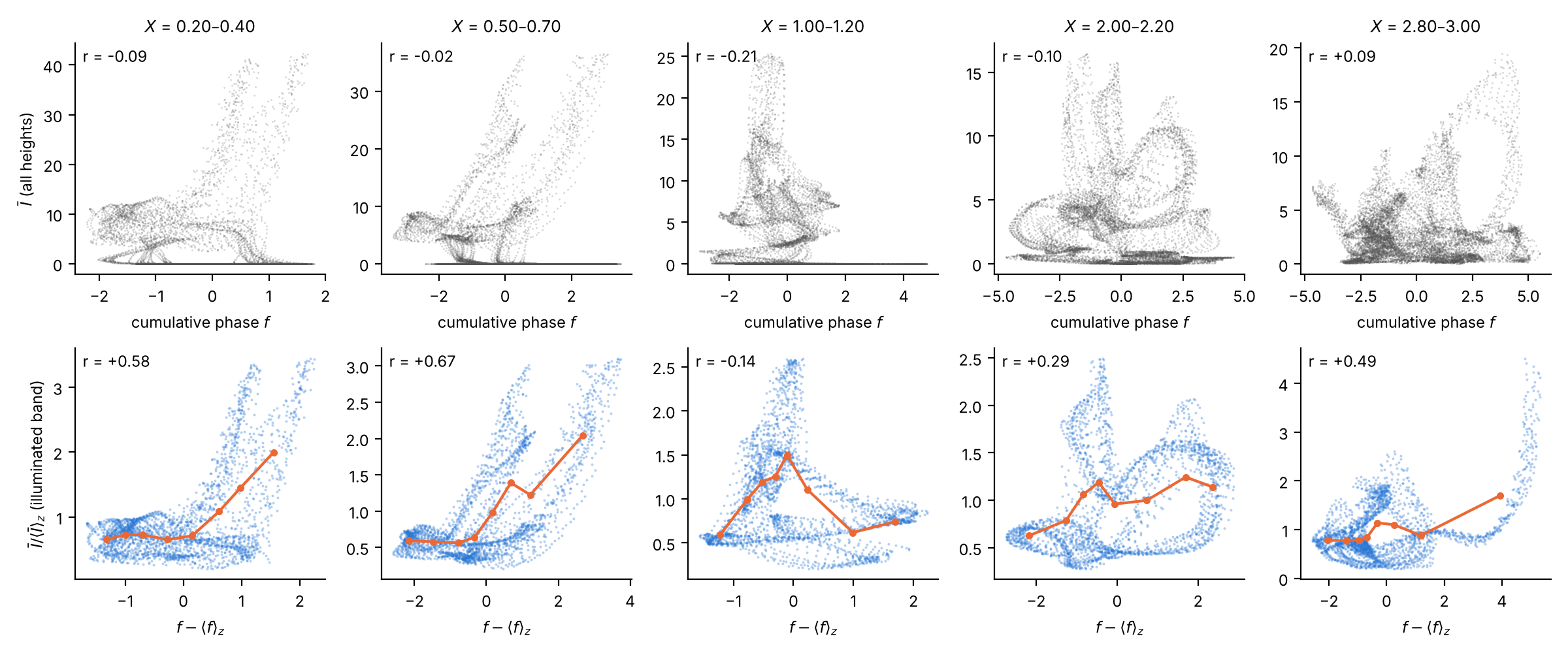}
\caption{Source-averaged intensity against cumulative phase in the simulations
of~\cite{US05} (case 2), for slabs of 21 successive ranges. Top: all heights, as
originally plotted. Bottom: heights within the illuminated band, with each range
normalised by its mean over $z$ and the mean of $f$ removed; the orange line is the
binned mean. The correlation coefficient of each panel is shown.}
\label{fig:scatter}
\end{figure}

A direct way to see how well the cumulative phase predicts the intensity is to plot
one against the other. We did this for the stored output of the simulations
of~\cite{US05} (one medium, 512 heights at 16 points per correlation length $L$, and
301 ranges at spacing $\Delta X=0.01$ in the scaled range $X=x/k_0L^2$, so that the
Fresnel distance is $X=1$ and the focusing range about $X\approx0.46$). The two cases
stored are independent media. Each panel of Figure~\ref{fig:scatter} pools a slab of
21 successive ranges ($\Delta X=0.2$).

Plotted over all heights (top row), the source-averaged intensity shows no trend with
the cumulative phase: the correlation lies between $-0.21$ and $+0.09$ for the five
slabs. This form of plot is, however, dominated by the many points outside the band
illuminated by the sources, where $\Ibar$ is close to zero whatever the phase. The
points also lie on curves and loops rather than in a cloud, because successive ranges
in a slab are strongly correlated: each slab traces a few dozen continuous
structures, not thousands of independent samples.

Restricting the plot to the illuminated band (120 heights, about 7.5 correlation
lengths) and normalising each range by its mean (bottom row) changes the picture only
near the source. For $X\lesssim0.7$ there is a clear trend, with correlation 0.58 and
0.67, but it is a threshold effect: the intensity is raised only where $f$ exceeds
its mean by more than about one standard deviation, and is roughly constant
elsewhere. Beyond the Fresnel distance the trend disappears. The binned mean is no
longer monotonic, and the correlations for the three slabs ($-0.14$, $+0.29$ and
$+0.49$) are carried by individual structures; the last, for example, by a single
large excursion of $f$. The per-range correlations~\eqref{eq:r}, averaged over
$1<X<2$ and $2<X<3$, are $-0.26$ and $+0.55$ for one medium and $+0.31$ and $+0.01$ for
the other, while the same correlations with the cumulative phase of the other
(independent) medium reach $\pm0.3$. With only two media they cannot be
distinguished from zero.

The ensemble results of Section~\ref{sec:cumphase} point the same way. Even where the
correlation is significant, the cumulative phase accounts for only part of the
variance of the source-averaged intensity: $r^2\approx0.35$--0.55 up to about $2x_f$,
falling to about 0.15 between $3x_f$ and $6x_f$. The cumulative phase therefore
identifies the strongest early ribbons, those above its largest excursions, but it does
not predict the intensity pattern in general, and it carries little information
beyond the Fresnel distance, over most of the range in which the ribbons
of~\cite{US05} were displayed. An animation of the bottom row as the slab moves through the range is available from
the authors.

\section{The fourth moment of the intensity}\label{sec:m4}

\subsection{Definitions}

For a single Gaussian beam or a plane wave, the normalised transverse covariance of the
intensity is
\begin{equation}\label{eq:CT}
  C_T(x,\zeta)=\frac{m_4(x,\zeta)}{\langle I(x,z)\rangle\,\langle I(x,z+\zeta)\rangle}-1,
  \qquad m_4(x,\zeta)=\langle I(x,z)\,I(x,z+\zeta)\rangle .
\end{equation}
Its value at $\zeta=0$ is the scintillation index, and we write $w(x)$ for its
half-width at half-maximum. Persistence along the range is measured by the
correlation coefficient at equal height,
\begin{equation}\label{eq:rhopar}
  \rho_\parallel(x_0;\xi)=\frac{\langle\delta I(x_0,z)\,\delta I(x_0+\xi,z)\rangle}
  {\bigl[\langle\delta I(x_0,z)^2\rangle\langle\delta I(x_0+\xi,z)^2\rangle\bigr]^{1/2}},
  \qquad \delta I=\frac{I}{\langle I\rangle}-1,
\end{equation}
and the persistence length $L_\parallel(x_0)$ is the separation at which
$\rho_\parallel=1/2$. For the plane wave, averages are taken over $z$ as well as over
media.

\subsection{Weak and saturated limits}

In the weak regime, \eqref{eq:jac} gives the covariance in terms of the fourth
derivative of the correlation function:
\begin{equation}\label{eq:CTweak}
  C_T(x,\zeta)\approx\mathrm{SI}(x)\,F(\zeta/l),\qquad
  F(u)=\Bigl(1-4u^2+\tfrac43u^4\Bigr)e^{-u^2}.
\end{equation}
The shape is independent of range, with $w=0.34\,l\approx1.01$, a first zero at
$0.52\,l\approx1.57$, and a negative lobe of depth $-0.62$ at $\zeta\approx2.9$. The
lobe is the statistical signature of a bright ribbon flanked by dark bands, from
which its energy was drawn.

In the saturated regime the field is a circular complex Gaussian, so
$C_T=|\Gamma_2(\zeta)|^2/\Gamma_2(0)^2$. For a plane wave the second moment is
$\Gamma_2=\exp(-\tfrac12D_f)$, where $D_f(\zeta)=2\sigma_f^2(1-e^{-\zeta^2/l^2})$ is
the structure function of the cumulative phase~\cite{Uscinski1977,Ishimaru1978}.
Hence
\begin{equation}\label{eq:CTsat}
  C_T(x,\zeta)\approx\exp\!\Bigl[-2\sigma_f^2\bigl(1-e^{-\zeta^2/l^2}\bigr)\Bigr],
  \qquad w_{\rm sat}=l\Bigl[-\ln\Bigl(1-\frac{\ln2}{2\sigma_f^2}\Bigr)\Bigr]^{1/2}
  \approx0.59\,\frac{l}{\sigma_f}\propto x^{-1/2},
\end{equation}
which is the coherence length of the field, the width of a speckle grain.

\subsection{Plane wave: height, width and persistence}

\begin{figure}[ht]
\centering
\includegraphics[width=\textwidth]{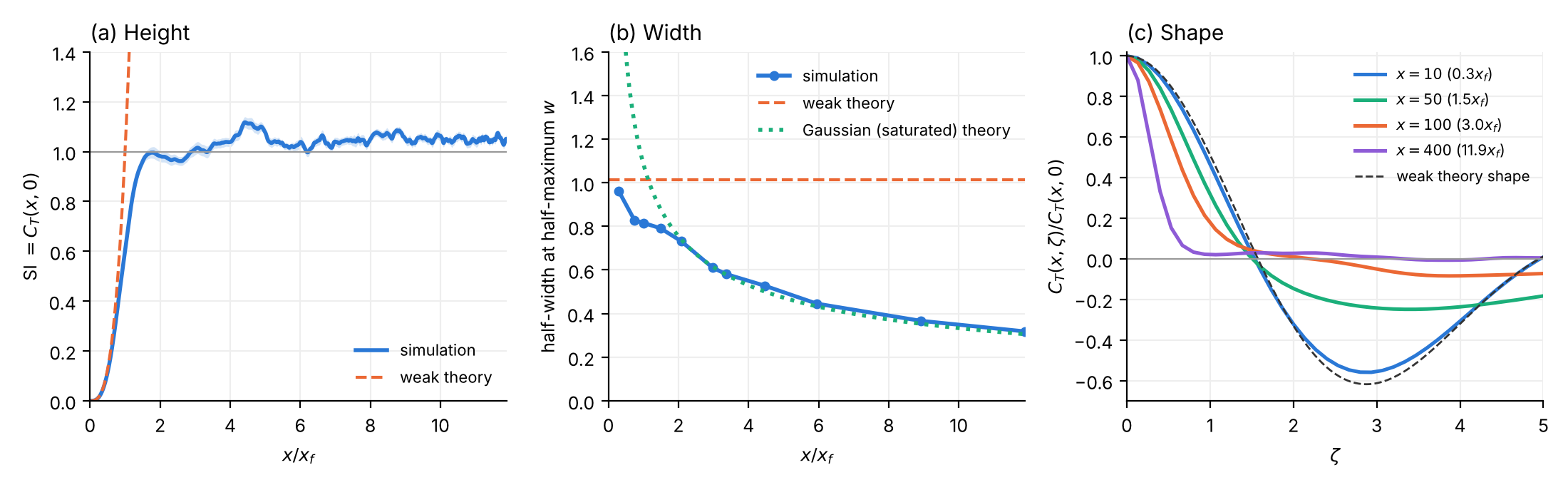}
\caption{Transverse covariance~\eqref{eq:CT} for a plane wave (50 media), with range
in units of $x_f$. (a) Height (the scintillation index), with the weak
estimate~\eqref{eq:SIweak}. (b) Half-width, with the weak shape~\eqref{eq:CTweak}
and the saturated form~\eqref{eq:CTsat}. (c) Normalised shapes at four ranges, with
the weak shape $F$.}
\label{fig:plane}
\end{figure}

\begin{table}[ht]
\centering
\caption{Plane-wave transverse covariance (50 media). Standard errors are at most
0.02 for SI and 0.01 for $w$.}
\label{tab:plane}
\small
\begin{tabular}{rrccccc}
\toprule
$x$ & $x/x_f$ & SI & $w$ & $w_{\rm sat}$, eq.~\eqref{eq:CTsat} & first zero & min$/$peak\\
\midrule
10  & 0.3  & 0.02 & 0.96 & --   & 1.54 & $-0.56$\\
25  & 0.7  & 0.33 & 0.83 & --   & 1.42 & $-0.37$\\
34  & 1.0  & 0.61 & 0.82 & --   & 1.43 & $-0.31$\\
50  & 1.5  & 0.96 & 0.79 & 0.88 & 1.51 & $-0.25$\\
70  & 2.1  & 0.98 & 0.73 & 0.74 & 1.71 & $-0.18$\\
100 & 3.0  & 1.01 & 0.61 & 0.62 & 2.21 & $-0.08$\\
150 & 4.5  & 1.12 & 0.53 & 0.50 & --   & $-0.06$\\
200 & 6.0  & 1.03 & 0.45 & 0.43 & --   & $-0.03$\\
300 & 8.9  & 1.06 & 0.37 & 0.35 & --   & $-0.02$\\
400 & 11.9 & 1.05 & 0.32 & 0.31 & --   & $-0.03$\\
\bottomrule
\end{tabular}
\end{table}

Figure~\ref{fig:plane} and Table~\ref{tab:plane} show two stages. Up to about $x_f$
the covariance has the weak form~\eqref{eq:CTweak}: at $x=10$ the half-width is 0.96
(theory 1.01), the first zero 1.54 (1.57) and the lobe $-0.56$ ($-0.62$). In this
range the peak describes ribbons of half-width about $l/3$, each flanked by dark
bands. Beyond about $2x_f$ the width follows the Gaussian prediction~\eqref{eq:CTsat}
to within 0.01--0.03 out to $x=400$, and the negative lobe disappears. The peak then
describes speckle grains, whose width shrinks as $x^{-1/2}$, rather than ribbons of
width comparable to $l$. What remains of ribbon-scale structure is small: an excess
$\mathrm{SI}-1$ of 0.03--0.12 and a low positive pedestal out to $\zeta\approx3$.

\begin{table}[ht]
\centering
\caption{Persistence along the range (plane wave, 50 media). The rms sideways shift is
that of the best-matching local pattern (windows of height $4l$, shifts up to
$\pm5$) between ranges $x_0$ and $x_0+10$; $\theta_{\rm rms}=\sqrt2\,\sigma_f/(k_0l)$ is
the rms ray slope.}
\label{tab:persist}
\small
\begin{tabular}{rccccc}
\toprule
$x_0$ & $L_\parallel$ & $w$ & $L_\parallel/k_0w^2$ & rms shift over 10 & $10\,\theta_{\rm rms}$\\
\midrule
25  & 19.9 & 0.83 & 2.31 & 0.58 & 0.54\\
50  & 17.3 & 0.79 & 2.20 & 0.67 & 0.77\\
100 & 10.4 & 0.61 & 2.21 & 1.35 & 1.09\\
200 & 5.9  & 0.45 & 2.36 & 1.90 & 1.54\\
\bottomrule
\end{tabular}
\end{table}

\begin{figure}[ht]
\centering
\includegraphics[width=0.9\textwidth]{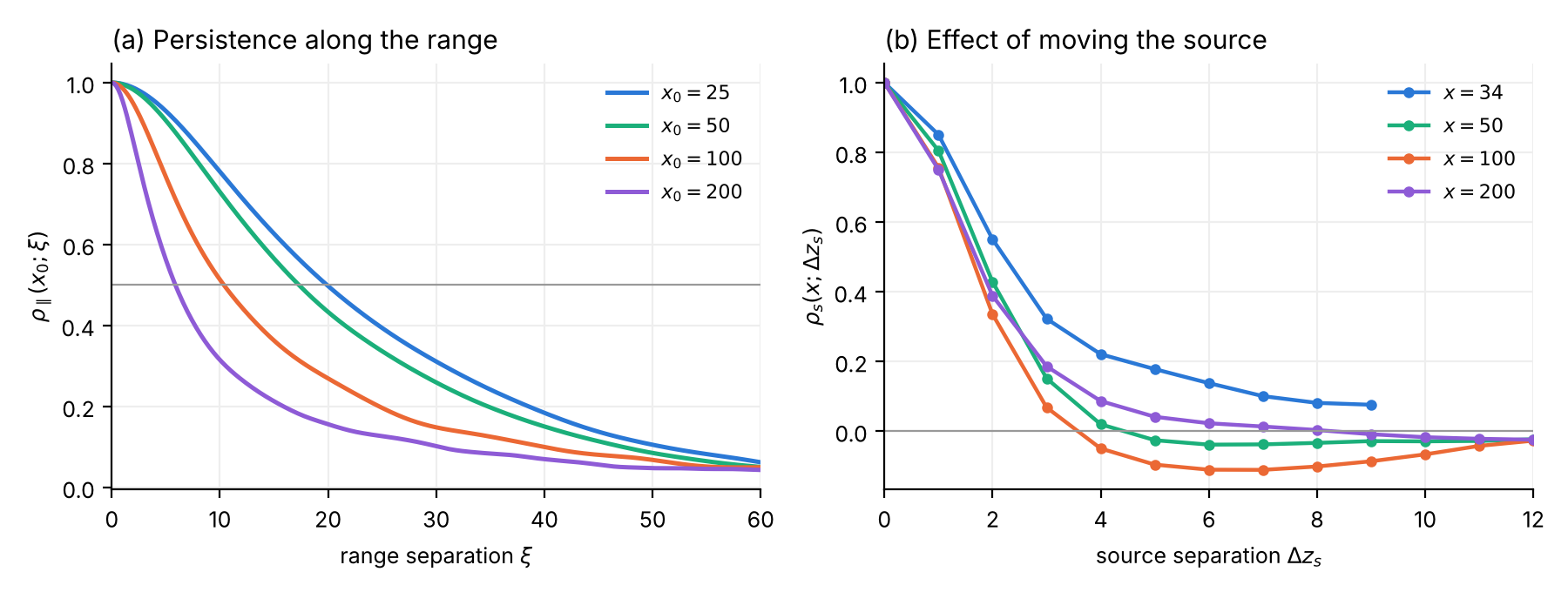}
\caption{(a) Correlation $\rho_\parallel$, equation~\eqref{eq:rhopar}, along the range
(plane wave). (b) Correlation $\rho_s$, equation~\eqref{eq:rhos}, between
single-source patterns whose sources are $\Delta z_s$ apart. For $x=34$ the beams
overlap only for $\Delta z_s\le9$. Both use 50 media.}
\label{fig:persist}
\end{figure}

Table~\ref{tab:persist} and Figure~\ref{fig:persist}(a) give the persistence. At every
range tested the persistence length is tied to the transverse width by
\begin{equation}\label{eq:Lpar}
  L_\parallel\approx2.3\,k_0w^2,
\end{equation}
the diffraction length of a structure of width $w$. Once the width is that of
speckle, the persistence length therefore falls as $1/x$, from 17 at $x_0=50$ to 6 at
$x_0=200$. Height and persistence of the plane-wave pattern are thus linked, but
beyond $2x_f$ both describe speckle rather than ribbons. Followed along the range, the
local pattern drifts sideways at roughly the rms ray slope (last two columns), so
individual structures slant by about the ray angle, consistent with the drifting
ribbons of Figure~\ref{fig:fields}.

\subsection{Source-averaged intensity}

The fourth moment computed in the project~\cite{Milir2025} was
\begin{equation}\label{eq:m4talk}
  m_4(x,\zeta)=\Bigl\langle\Ibar(x,0)\,\tfrac12\bigl[\Ibar(x,\zeta)+\Ibar(x,-\zeta)\bigr]\Bigr\rangle ,
\end{equation}
for the source-averaged intensity, with the reference fixed at $z=0$, over 50
realisations, at the single range $x=50$. Because it is neither normalised nor
mean-subtracted, it splits as
\begin{equation}\label{eq:m4split}
  m_4(x,\zeta)=\langle\Ibar(x,0)\rangle\,\langle\Ibar(x,\zeta)\rangle_\pm\,
  \bigl[1+C_{\Ibar}(x,\zeta)\bigr],
\end{equation}
where $\langle\cdot\rangle_\pm$ is the mean of the values at $\pm\zeta$. The first
factor is the product of the mean intensities, which depends only on the spread of the
source positions and the beams; the fluctuation part $C_{\Ibar}$ contains the medium.
We computed~\eqref{eq:m4talk} at ranges chosen in units of $x_f$ and $x_F$, and
estimated $C_{\Ibar}$ both with the reference fixed at $z=0$ and with the reference
averaged over $|z_0|\le5$, where the mean envelope is flat. The two estimates agree
within their standard errors, the latter being somewhat less noisy.

\begin{figure}[ht]
\centering
\includegraphics[width=\textwidth]{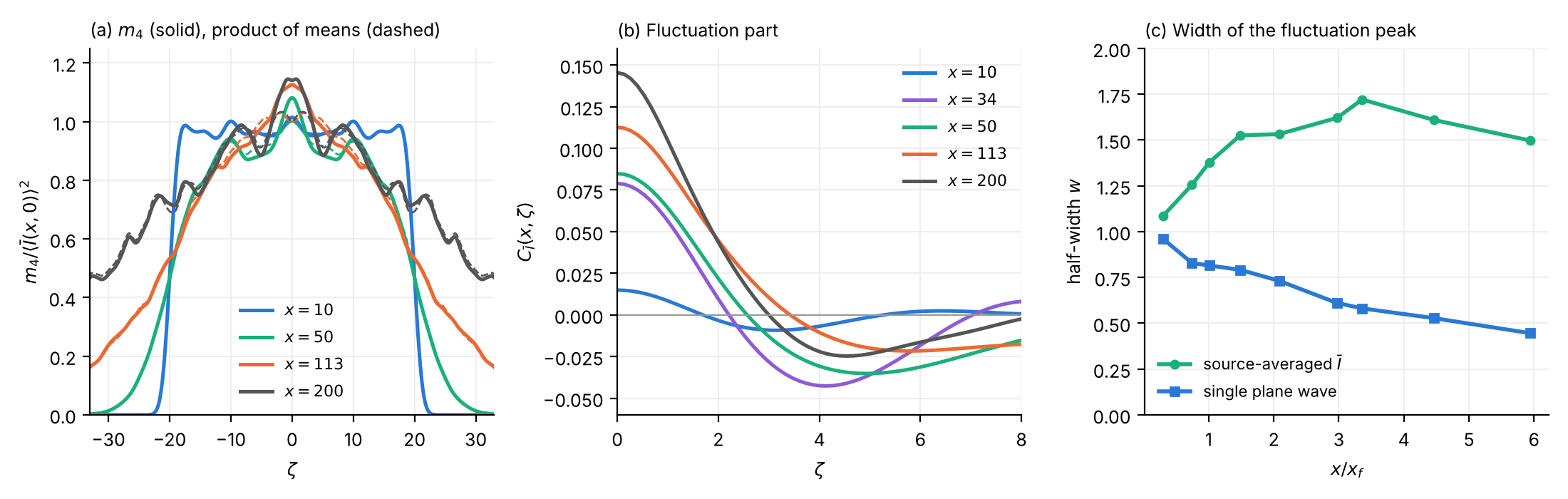}
\caption{Fourth moment of the source-averaged intensity (50 media, 40 sources).
(a) The moment~\eqref{eq:m4talk}, normalised by $\langle\Ibar(x,0)\rangle^2$
(solid), and the product of mean intensities (dashed). (b) The fluctuation part
$C_{\Ibar}$ of~\eqref{eq:m4split}, reference averaged over $|z_0|\le5$. (c)
Half-width of the fluctuation peak, for $\Ibar$ and for a single plane wave.}
\label{fig:srcavg}
\end{figure}

\begin{table}[ht]
\centering
\caption{Fluctuation part of the fourth moment of the source-averaged intensity (50
media). The fixed-reference estimate is given $\pm$ its bootstrap standard error.}
\label{tab:srcavg}
\small
\begin{tabular}{rcccccc}
\toprule
$x$ & $x/x_f$ & $x/x_F$ & $C_{\Ibar}(x,0)$, $z_0=0$ & $C_{\Ibar}(x,0)$, $|z_0|\le5$ & $w$ & lobe minimum\\
\midrule
10  & 0.3 & 0.09 & $0.015\pm0.003$ & 0.015 & 1.09 & $-0.009$\\
25  & 0.7 & 0.22 & $0.058\pm0.009$ & 0.072 & 1.26 & $-0.038$\\
34  & 1.0 & 0.30 & $0.059\pm0.010$ & 0.079 & 1.38 & $-0.043$\\
50  & 1.5 & 0.44 & $0.081\pm0.012$ & 0.085 & 1.53 & $-0.035$\\
70  & 2.1 & 0.62 & $0.070\pm0.013$ & 0.079 & 1.53 & $-0.023$\\
100 & 3.0 & 0.88 & $0.094\pm0.017$ & 0.095 & 1.62 & $-0.018$\\
113 & 3.4 & 1.00 & $0.126\pm0.025$ & 0.113 & 1.72 & $-0.022$\\
150 & 4.5 & 1.33 & $0.125\pm0.028$ & 0.120 & 1.61 & $-0.015$\\
200 & 6.0 & 1.77 & $0.141\pm0.025$ & 0.145 & 1.50 & $-0.025$\\
\bottomrule
\end{tabular}
\end{table}

Figure~\ref{fig:srcavg} and Table~\ref{tab:srcavg} show the results. The unnormalised
moment~\eqref{eq:m4talk} is almost entirely the product of the mean intensities: its
flat top and its fall near $|\zeta|\approx20$ reflect the span of source positions, and
the medium contributes a bump of relative height only about 0.08 at $x=50$. The
fluctuation part rises from 0.015 at $x=10$ to about 0.08 by $x_f$ and to 0.145 at
$x=200$. Its half-width, however, rises to about 1.5--1.7, roughly $l/2$, by
$1.5x_f$ and then stays there, whereas the plane-wave width keeps narrowing as speckle
develops (Figure~\ref{fig:srcavg}(c)). A shallow negative lobe near $\zeta\approx4$--5,
deepest ($-0.04$) around $x_f$, marks the dark bands beside the averaged ribbons.
Averaging over 40 sources thus removes most of the source-dependent speckle and keeps
structure of ribbon width: the width of $C_{\Ibar}$ measures the transverse size of
the ribbons, and its height their contrast after averaging.

\subsection{Moving the source}

To measure directly how much of the pattern survives a change of source position, we
correlate the normalised fluctuations of single-source patterns whose sources are
$\Delta z_s$ apart:
\begin{equation}\label{eq:rhos}
  \rho_s(x;\Delta z_s)=\frac{\sum\delta_s\,\delta_{s'}}{\bigl[\sum\delta_s^2\sum\delta_{s'}^2\bigr]^{1/2}},
  \qquad \delta_s=\frac{I(x,z;z_s)}{\langle I(x,z;z_s)\rangle}-1,\quad z_{s'}=z_s+\Delta z_s,
\end{equation}
the sums running over media, over pairs of sources, and over the heights where both
mean beam profiles exceed 30\% of their maxima. Figure~\ref{fig:persist}(b) shows that
for $x\ge50$ the correlation falls to $1/2$ at $\Delta z_s\approx1.6$--1.8, about
$0.55\,l$, and is close to zero by $\Delta z_s\approx3$--4. A small positive
residual, $\rho_s\approx0.05$--0.09 at $\Delta z_s=4$, appears only for $x\ge150$. At
$x=34$ the decorrelation is slower (half at 2.2) and a residual of 0.2 persists to
$\Delta z_s=4$.

One complication applies to single beams. Their rms sideways wander,
$\theta_{\rm rms}x$, is comparable with the beam width beyond $x\approx70$, so the
ensemble-mean beam profile is broadened, and the normalised index in the beam core
rises to 1.5--2.0 for $x\ge100$. Single-beam statistics therefore mix wander with
speckle, which is why the plane wave is used for the basic statistics above.

\section{Discussion}\label{sec:discussion}

\paragraph{Formation of ribbons.} Near the source the ribbons are well predicted by
the medium: the source-averaged intensity correlates at 0.84--0.88 with the curvature
of the cumulative phase, in line with the geometric-optics relation~\eqref{eq:jac},
and the transverse fourth moment has the weak-scattering shape, with ribbons of
half-width about $l/3$ flanked by dark bands. The cumulative phase itself is a weaker
predictor in this range. The picture of~\cite{US05}, in which a large excursion of $f$
acts as a converging lens, captures the mechanism only loosely, since the curvature,
not the excursion, is what focuses, and the correlation between the two has 
magnitude only $1/\sqrt3$.

\paragraph{The straight-line picture beyond the focusing range.} Both $f$ and
relation~\eqref{eq:jac} integrate the medium along straight horizontal lines. That is
justified only while transverse ray displacements remain small compared with $l$,
that is, before the first caustics. Beyond $x_f$ the pattern drifts sideways at about
the rms ray slope, and the ribbons slant and branch (Figure~\ref{fig:fields}), while
$f$ remains horizontal by construction. The slow decline of the correlation between
$\Ibar$ and $f$ beyond $x_f$ may therefore partly reflect the persistence of $f$ itself
rather than a causal link. In the branched-flow literature the branches are traced to
caustics of the ray flow rather than to valleys of the potential~\cite{Kaplan2002}.
The scatter plots of Section~\ref{sec:scatter} show this directly for the original
simulations: beyond the Fresnel distance no consistent relation between the
cumulative phase and the intensity can be found. A predictor that integrates along
the actual ray paths, or the range-weighted curvature of~\eqref{eq:jac}, would be a
natural test.

\paragraph{Persistence.} For a plane wave the fourth moment shows a precise link
between transverse size and persistence, $L_\parallel\approx2.3\,k_0w^2$, but beyond
about $2x_f$ both are those of Gaussian speckle. The persistence of ribbons is
therefore a property of the source-averaged intensity, whose correlation width stays
near $l/2$ while the plane-wave width shrinks. Moving the source by about $0.55\,l$
halves the correlation of the single-source pattern, and the part that is independent
of the source position is a few per cent of the fluctuation variance in the ranges
studied here.

\paragraph{Dark regions.} The dark regions change character with the regime. Before
$x_f$ they are defocused regions, where the curvature of $f$ has the diverging sign;
the conditional means show that they are tied to minima of $f$ as strongly as bright
regions are tied to maxima, and the negative lobe of the fourth moment shows them as
bands beside each ribbon. Around focusing they are increasingly shadow zones between
pairs of caustics. In saturation, low intensity is controlled by the zeros of an
approximately Gaussian complex field. In the $(x,z)$ plane these are isolated points
(wave dislocations), around which the phase winds by $\pm2\pi$ and the intensity rises
quadratically~\cite{NyeBerry1974}. Their density follows from the spectrum of the
field~\cite{BerryDennis2000}, they carry topological charge, and as the source moves
they move continuously and are created or annihilated only in pairs. A persistent
dark region in the source-averaged intensity, which itself has no zeros, corresponds
to the zeros of the individual fields remaining clustered as the source moves. This
suggests a statistical description of the dark regions through the density, core
shapes and trajectories of the zeros, which we leave for future work.

\paragraph{Discrete models.} The project~\cite{BMJ} also explored discrete,
Markov-chain models in which ``particles'' step transversely with probabilities
derived from the local medium. These are not pursued here; this discrete model and
its connection to the continuum model in different parameter ranges will be 
explored in a further publication.

\section{Conclusions}\label{sec:conclusions}

We have revisited the explanation of high-intensity ribbons in random media in terms
of the cumulative phase of the medium, using a grid-independent continuous medium,
ensembles of 50 media and explicitly defined statistics. The main findings are as
follows.
\begin{itemize}
\item The focusing range $x_f\approx l(4\sqrt\pi\mu^2)^{-1/3}$ organises the results.
  Up to $x_f$, ribbons are well predicted by the transverse curvature of the cumulative
  phase (correlation 0.84--0.88), and the fourth moment has the weak-scattering form.
\item The cumulative phase itself correlates more weakly near the source but more
  persistently, averaging about 0.4 between $3x_f$ and $6x_f$; part of this persistence is a consequence
  of its construction as a running sum. Even so, it accounts for at most about half of
  the variance of the source-averaged intensity, and in the original simulations no
  consistent relation is found beyond the Fresnel distance.
\item Beyond about $2x_f$ the plane-wave fourth moment is that of Gaussian speckle,
  with width $\propto x^{-1/2}$ and persistence length $\approx2.3\,k_0w^2$. Ribbon-scale
  structure survives in the source-averaged intensity, whose correlation width remains
  close to $l/2$.
\item Bright and dark regions are associated symmetrically with maxima and minima of
  the cumulative phase up to about $3x_f$. A description of the persistent dark regions
  through the zeros of the field is suggested.
\end{itemize}

\section*{Acknowledgements}
This work began as a summer research project supported by the Faculty of Mathematics,
University of Cambridge, the Cantab Capital Institute for the Mathematics of
Information (CCIMI) and G-Research.

\end{document}